\documentclass{article}
\usepackage{amsmath}

\numberwithin{equation}{section}
\input{tcilatex}
\begin{document}


\centerline{\large\bf  The $n$-tangle of odd $n$ qubits\footnote{The paper
was supported by NSFC (Grant No. 10875061) and Tsinghua National Laboratory for Information Science and
Technology. }} 

\vspace*{8pt}
\centerline{Dafa Li\footnote{email
address:dli@math.tsinghua.edu.cn}}

\centerline{ Dept of mathematical sciences}

\centerline{Tsinghua University, Beijing 100084 CHINA}

\abstract
Coffman, Kundu and Wootters presented the 3-tangle of three qubits in [Phys.
Rev. A 61, 052306 (2000)]. Wong and Christensen extended the 3-tangle to
even number of qubits, known as $n$-tangle [Phys. Rev. A 63, 044301 (2001)].
In this paper, we propose a generalization of the 3-tangle to any odd $n$%
-qubit pure states and call it the $n$-tangle of odd $n$ qubits. We show
that the $n$-tangle of odd $n$ qubits is invariant under permutations of the
qubits, and is an entanglement monotone. The $n$-tangle of odd $n$ qubits
can be considered as a natural entanglement measure of any odd $n$-qubit
pure states.

Keywords: 3-tangle, $n$-tangle of odd $n$ qubits, concurrence, residual entanglement

PACS numbers: 03.67.Mn, 03.65.Ud

\section{Introduction}

Quantum entanglement is a key quantum mechanical resource in quantum
computation and information, such as quantum cryptography, quantum dense
coding and quantum teleportation \cite{Horodecki}. Entanglement measure,
which characterizes the degree of entanglement contained in a quantum state,
has been a subject under intensive research.

The entanglement of bipartite systems is well understood. The concurrence 
\cite{Wootters} is a good entanglement measure for two-qubit states and is
an entanglement monotone, i.e., it is non-increasing under local quantum
operations and classical communication (LOCC). Generalizations of the
concurrence to higher dimensions can be found, for example, in \cite%
{Rungta,Uhlmann}. The residual entanglement, or the 3-tangle has been
constructed in terms of the concurrences as a widely accepted entanglement
measure to quantify the entanglement in three-qubit pure states \cite%
{Coffman}. The 3-tangle is permutationally invariant, is an entanglement
monotone, and is a SLOCC (stochastic local operations and classical
communication) polynomial of degree 4. Furthermore, the 3-tangle is bounded
between 0 and 1, and it assumes value 1 for the GHZ state and vanishes for
the W state \cite{Coffman,Dur}. Several other measures have been constructed
specifically for the entanglement of the three-qubit pure states \cite%
{Lee,Emary,Ou}. The partial tangle, reported in \cite{Lee}, represents the
residual two-qubit entanglement of a three-qubit pure state and reduces to
the two-qubit concurrence for the W state. The $\sigma $-measure \cite{Emary}
and $\pi $-tangle \cite{Ou} have been introduced as entanglement monotones
for genuine three-qubit entanglement. Whereas the 3-tangle vanishes for the
W state, both $\sigma $-measure and $\pi $-tangle take non-zero values for
the W state as well as the GHZ state. Many other entanglement measures for
quantifying the entanglement of multipartite pure states have been proposed 
\cite{Miyake,Wong,Sharma,LDF07a,LDF09b} (see also the review \cite{Horodecki}
and references therein). Hyperdeterminant, as a generalization of the
concurrence and the 3-tangle, has been shown to be an entanglement monotone
and describes the genuine multipartite entanglement \cite{Miyake}. The $n$%
-tangle is a straightforward extension of 3-tangle to even number of qubits 
\cite{Wong}. As has been previously noted, the $n$-tangle is the square of
generalization of the concurrence, is invariant under permutations, and is
an entanglement monotone. Like the 3-tangle, the $n$-tangle is equal to 1
for the GHZ state and vanishes for the W state \cite{Wong}. However the $n$%
-tangle is not residual entanglement for four or more qubits \cite{LDFQIC10}%
. It has been found that the 4-tangle for four-qubit states can be
interpreted as a type of residual entanglement similar to the interpretation
of 3-tangle for three-qubit states as the residual tangle \cite{Gour}. An
alternative 4-tangle has recently been obtained by using negativity fonts
and the 4-tangle is a genuine entanglement measure of four-qubit pure states 
\cite{Sharma}. In \cite{LDF07a}, the residual entanglement of odd $n$ qubits
has been proposed as an entanglement measure for odd $n$-qubit pure states
and shown to be an entanglement monotone \cite{LDF09b}. The odd $n$-tangle
(although called odd $n$-tangle, it is not defined in the same way as has
been done for the $n$-tangle by directly extending the definition of
3-tangle to even $n$ qubits) has been defined by taking the average of the
residual entanglement with respect to qubit $i$, which is obtained from the
residual entanglement of odd $n$ qubits under transposition on qubits 1 and $%
i$ \cite{LDF09b}. It has been shown that the odd $n$-tangle is
permutationally invariant, $SL$-invariant and $LU$-invariant, and is an
entanglement monotone \cite{LDF09b}.

In this paper, we give an alternative formulation of the 3-tangle. We extend
the formulation in a straightforward way to any odd $n$-qubit pure states
and define the $n$-tangle with respect to qubit $i$. By taking the average
of the $n$-tangle with respect to qubit $i$, we define the $n$-tangle of odd 
$n$ qubits, which is invariant under permutations of the qubits. The
extended formulation is then reduced by using simple mathematics. It turns
out that the $n$-tangle with respect to qubit $i$ and the $n$-tangle of odd $%
n$ qubits are equal to the residual entanglement with respect to qubit $i$
and the odd $n$-tangle respectively, and consequently the former inherit the
properties of the latter, like the monotonicity, invariance under $SL$ and $%
LU$ operations as well as the property of satisfying SLOCC equation.
Moreover, the $n$-tangle with respect to qubit $i$ is a SLOCC polynomial of
degree 4. Like the 3-tangle, the $n$-tangle of odd $n$ qubits takes value 1
for the GHZ state and vanishes for the W state. Finally we extend the $n$%
-tangle of odd $n$ qubits to mixed states via the convex roof construction.
This work will extend our understanding of multipartite entanglement.

The rest of the paper is organized as follows. In Section 2 we briefly
review the definitions and the formulations of the concurrence, the 3-tangle
and the $n$-tangle. We then give an alternative formulation of the 3-tangle
and extend it to odd $n$ qubits. We also introduce the definitions of the $n$%
-tangle with respect to qubit $i$ and the $n$-tangle of odd $n$ qubits. In
Section 3, we study the $n$-tangle with respect to qubit $i$ and the $n$%
-tangle of odd $n$ qubits in more detail and we discuss their properties.
Finally, we draw our conclusion in Section 4.

\section{The $n$-tangle of odd $n$ qubits}

\subsection{Preliminaries}

The concurrence for two-qubit pure states is defined as $C(\psi )=\left\vert
\langle \psi \right\vert \tilde{\psi}\rangle |^{2}$ \cite{Wootters}, where $|%
\tilde{\psi}\rangle $ denotes the resulting state after applying the
operator $\sigma _{y}\otimes \sigma _{y}$ to the complex conjugate of $|\psi
\rangle $ \cite{Wootters}, i.e. $|\tilde{\psi}\rangle =\sigma _{y}\otimes
\sigma _{y}|\psi ^{\ast }\rangle $. Here the asterisk indicates complex
conjugatation in the standard basis. For three-qubit pure states, the
3-tangle $\tau _{ABC}$ (or $\tau _{123}$)\ can be calculated by means of
concurrences and is given by $\tau
_{ABC}=C_{A(BC)}^{2}-C_{AB}^{2}-C_{AC}^{2} $ \cite{Coffman}, where $C_{AB}$
and $C_{AC}$ are the concurrences of the corresponding two-qubit subsytems $%
\rho _{AB}$ and $\rho _{AC}$, respectively, and $C_{A(BC)}^{2}=4\det \rho
_{A}$. Here $\rho _{AB}$, $\rho _{AC}$ and $\rho _{A}$ are the reduced
density matrices. Let $|\psi \rangle =\sum_{i=0}^{7}a_{i}|i\rangle $, where $%
\sum_{i=0}^{7}|a_{i}|=1$. An expression of the 3-tangle in terms of the
coefficients for the state $|\psi \rangle $ is given by \cite{Coffman} 
\begin{equation}
\tau _{123}=4\bigl|d_{1}-2d_{2}+4d_{3}\bigr|,
\end{equation}%
where 
\begin{align}
d_{1}&
=a_{0}^{2}a_{7}^{2}+a_{1}^{2}a_{6}^{2}+a_{2}^{2}a_{5}^{2}+a_{3}^{2}a_{4}^{2},
\\
d_{2}&
=a_{0}a_{7}a_{3}a_{4}+a_{0}a_{7}a_{2}a_{5}+a_{0}a_{7}a_{1}a_{6}+a_{3}a_{4}a_{2}a_{5}+a_{3}a_{4}a_{1}a_{6}+a_{2}a_{5}a_{1}a_{6},
\\
d_{3}& =a_{0}a_{6}a_{5}a_{3}+a_{7}a_{1}a_{2}a_{4}.
\end{align}%
A more standard form of the 3-tangle is given as follows \cite{Coffman}: 
\begin{equation}
\tau _{123}=2\Bigl|\sum a_{\alpha _{1}\alpha _{2}\alpha _{3}}a_{\beta
_{1}\beta _{2}\beta _{3}}a_{\gamma _{1}\gamma _{2}\gamma _{3}}a_{\delta
_{1}\delta _{2}\delta _{3}}\times \epsilon _{\alpha _{1}\beta _{1}}\epsilon
_{\alpha _{2}\beta _{2}}\epsilon _{\gamma _{1}\delta _{1}}\epsilon _{\gamma
_{2}\delta _{2}}\epsilon _{\alpha _{3}\gamma _{3}}\epsilon _{\beta
_{3}\delta _{3}}\Bigr|,  \label{3-tangle}
\end{equation}%
where the sum is over all the indices, $\alpha _{l}$, $\beta _{l}$, $\gamma
_{l}$, and $\delta _{l}$\ $\in \{0,1\}$, $\epsilon _{00}=\epsilon _{11}=0$,
and $\epsilon _{01}=-\epsilon _{10}=1$. The above formulation of the
3-tangle is invariant under permutations of the qubits.

Let $|\psi \rangle $\ be any state of $n$ qubits and $|\psi \rangle
=\sum_{i=0}^{2^{n}-1}a_{i}|i\rangle $, where $\sum_{i=0}^{2^{n}-1}|a_{i}|=1$%
.\ The $n$-tangle is defined for the state $|\psi \rangle $ as follows \cite%
{Wong}: 
\begin{align}
\tau _{12\cdots n}& =2\Bigl|\sum a_{\alpha _{1}\cdots \alpha _{n}}a_{\beta
_{1}\cdots \beta _{n}}a_{\gamma _{1}\cdots \gamma _{n}}a_{\delta _{1}\cdots
\delta _{n}}  \notag \\
& \quad \times \epsilon _{\alpha _{1}\beta _{1}}\epsilon _{\alpha _{2}\beta
_{2}}\cdots \epsilon _{\alpha _{n-1}\beta _{n-1}}\epsilon _{\gamma
_{1}\delta _{1}}\epsilon _{\gamma _{2}\delta _{2}}\cdots \epsilon _{\gamma
_{n-1}\delta _{n-1}}\epsilon _{\alpha _{n}\gamma _{n}}\epsilon _{\beta
_{n}\delta _{n}}\Bigr|,  \label{odd-tangle-wong}
\end{align}%
for all even $n$ and $n=3$. However the above formula is not invariant under
permutations of qubits for odd $n>3$, and therefore, the $n$-tangle remains
undefined for odd $n>3$ \cite{Wong}.

\subsection{Alternative formulation of the 3-tangle}

Here we let 
\begin{align}
\tau _{123}^{(1)}& =2\Bigl|\sum a_{\alpha _{1}\alpha _{2}\alpha
_{3}}a_{\beta _{1}\beta _{2}\beta _{3}}a_{\gamma _{1}\gamma _{2}\gamma
_{3}}a_{\delta _{1}\delta _{2}\delta _{3}}\times \epsilon _{\alpha _{2}\beta
_{2}}\epsilon _{\alpha _{3}\beta _{3}}\epsilon _{\gamma _{2}\delta
_{2}}\epsilon _{\gamma _{3}\delta _{3}}\epsilon _{\alpha _{1}\gamma
_{1}}\epsilon _{\beta _{1}\delta _{1}}\Bigr|,  \label{3-tangle-1} \\
\tau _{123}^{(2)}& =2\Bigl|\sum a_{\alpha _{1}\alpha _{2}\alpha
_{3}}a_{\beta _{1}\beta _{2}\beta _{3}}a_{\gamma _{1}\gamma _{2}\gamma
_{3}}a_{\delta _{1}\delta _{2}\delta _{3}}\times \epsilon _{\alpha _{1}\beta
_{1}}\epsilon _{\alpha _{3}\beta _{3}}\epsilon _{\gamma _{1}\delta
_{1}}\epsilon _{\gamma _{3}\delta _{3}}\epsilon _{\alpha _{2}\gamma
_{2}}\epsilon _{\beta _{2}\delta _{2}}\Bigr|,  \label{3-tangl-2} \\
\tau _{123}^{(3)}& =2\Bigl|\sum a_{\alpha _{1}\alpha _{2}\alpha
_{3}}a_{\beta _{1}\beta _{2}\beta _{3}}a_{\gamma _{1}\gamma _{2}\gamma
_{3}}a_{\delta _{1}\delta _{2}\delta _{3}}\times \epsilon _{\alpha _{1}\beta
_{1}}\epsilon _{\alpha _{2}\beta _{2}}\epsilon _{\gamma _{1}\delta
_{1}}\epsilon _{\gamma _{2}\delta _{2}}\epsilon _{\alpha _{3}\gamma
_{3}}\epsilon _{\beta _{3}\delta _{3}}\Bigr|.  \label{3-tangle-3}
\end{align}%
Inspection of Eqs. (\ref{3-tangle}) and (\ref{3-tangle-3}) reveals that $%
\tau _{123}=\tau _{123}^{(3)}$. Indeed, a direct calculation gives $\tau
_{123}^{(1)}=\tau _{123}^{(2)}=\tau _{123}^{(3)}$. Now, let us look at the
formulas from a different perspective. We note that $\tau _{123}^{(2)}$ can
be obtained from $\tau _{123}^{(1)}$ by taking the transposition $(1,2)$ on
qubits $1$ and $2$. Analogously, $\tau _{123}^{(3)}$ can be obtained from $%
\tau _{123}^{(1)}$ by taking the transposition $(1,3)$ on qubits $1$ and $3$%
. It turns out that we can also obtain $\tau _{123}^{(1)}=\tau
_{123}^{(2)}=\tau _{123}^{(3)}$ by using the fact that the 3-tangle $\tau
_{123}$ is invariant under permutations of the three qubits \cite{Coffman}.
We may thus rewrite the 3-tangle as follows: 
\begin{equation}
\tau _{123}=(\tau _{123}^{(1)}+\tau _{123}^{(2)}+\tau _{123}^{(3)})/3.
\label{new-3-tangle}
\end{equation}

\subsection{The $n$-tangle with respect to qubit $i$ and the $n$-tangle of
odd $n$ qubits}

We extend Eqs. (\ref{3-tangle-1})-(\ref{3-tangle-3}) to any odd $n$ qubits.
Let 
\begin{align}
\tau _{12\cdots n}^{(i)}& =2\bigl|W_{12\cdots n}^{(i)}\bigr|,
\label{odd-tangle-1} \\
W_{12\cdots n}^{(i)}& =\sum a_{\alpha _{1}\cdots \alpha _{n}}a_{\beta
_{1}\cdots \beta _{n}}a_{\gamma _{1}\cdots \gamma _{n}}a_{\delta _{1}\cdots
\delta _{n}}\times \epsilon _{\alpha _{i}\gamma _{i}}\epsilon _{\beta
_{i}\delta _{i}}  \notag \\
& \quad \times \epsilon _{\alpha _{1}\beta _{1}}\cdots \epsilon _{\alpha
_{i-1}\beta _{i-1}}\epsilon _{\alpha _{i+1}\beta _{i+1}}\cdots \epsilon
_{\alpha _{n}\beta _{n}}  \notag \\
& \quad \times \epsilon _{\gamma _{1}\delta _{1}}\cdots \epsilon _{\gamma
_{i-1}\delta _{i-1}}\epsilon _{\gamma _{i+1}\delta _{i+1}}\cdots \epsilon
_{\gamma _{n}\delta _{n}},  \label{odd-tangle-2}
\end{align}%
where the sum is over all the indices and $i=1$, $\cdots $, $n$. One can
verify that $\tau _{12\cdots n}^{(i)}$ with $n\geq 5$ is invariant under any
permutation of all but qubit $i$. So, we call $\tau _{12\cdots n}^{(i)}$ the 
$n$-tangle with respect to qubit $i$. One can show that $\tau _{12\cdots
n}^{(1)}$ turns into $\tau _{12\cdots n}^{(i)}$ under the transposition $%
(1,i)$ on qubits 1 and $i$, $i=2,3,\cdots ,n$.

In analogy to Eq. (\ref{new-3-tangle}), we define the $n$-tangle of odd $n$
qubits as follows: 
\begin{equation}
\tau _{12\cdots n}=\frac{1}{n}\sum_{i=1}^{n}\tau _{12\cdots n}^{(i)}.
\label{odd-tangle-3}
\end{equation}%
It is not hard to see that $\tau _{12\cdots n}$ is invariant under all the
permutations of the qubits, and the values of $\tau _{12\cdots n}^{(i)}$ and 
$\tau _{12\cdots n}$ are bounded between $0$ and $1$. Note also that when $%
n=3$, $\tau _{12\cdots n}^{(i)}$ and $\tau _{12\cdots n}$ become $\tau
_{123} $.

\subsection{Reduction of the formulation}

We observe that it takes $3\cdot 2^{4n}$ multiplications to compute $\tau
_{12\cdots n}^{(i)}$ by Eqs. (\ref{odd-tangle-1}) and (\ref{odd-tangle-2}).
Next we reduce the formulation of $\tau _{12\cdots n}^{(1)}$. From Eq. (\ref%
{odd-tangle-2}), we have 
\begin{align}
W_{12\cdots n}^{(1)}& =\sum a_{\alpha _{1}\cdots \alpha _{n}}a_{\beta
_{1}\cdots \beta _{n}}a_{\gamma _{1}\cdots \gamma _{n}}a_{\delta _{1}\cdots
\delta _{n}}\times \epsilon _{\alpha _{1}\gamma _{1}}\epsilon _{\beta
_{1}\delta _{1}}  \notag \\
& \quad \times \epsilon _{\alpha _{2}\beta _{2}}\cdots \epsilon _{\alpha
_{n}\beta _{n}}\epsilon _{\gamma _{2}\delta _{2}}\cdots \epsilon _{\gamma
_{n}\delta _{n}}.  \label{odd-n-tangle-3}
\end{align}%
After some calculations, we obtain (we refer the reader to Appendix A for
details) 
\begin{align}
W_{12\cdots n}^{(1)}& =2(PQ-T^{2})\text{,}  \label{reduce-1} \\
\tau _{12\cdots n}^{(1)}& =4\bigl|T^{2}-PQ\bigr|,  \label{n-tangle-8}
\end{align}%
where 
\begin{align}
T& =\sum_{i=0}^{2^{n-1}-1}(-1)^{N(i)}a_{i}a_{2^{n}-i-1},  \label{n-tangle-9}
\\
P& =2\sum_{i=0}^{2^{n-2}-1}(-1)^{N(i)}a_{2i}a_{2^{n-1}-2i-1},
\label{n-tangle-10} \\
Q& =2\sum_{i=0}^{2^{n-2}-1}(-1)^{N(i)}a_{2^{n-1}+2i}a_{2^{n}-2i-1}.
\label{n-tangle-11}
\end{align}%
Here $N(l)$ is the number of 1s in the $n$-bit binary representation $%
l_{n-1}...l_{1}l_{0}$ of $l$. We further note that it takes $(2^{n}+3)$
multiplications to compute $\tau _{12\cdots n}^{(1)}$ using Eqs. (\ref%
{n-tangle-8})-(\ref{n-tangle-11}). A plain calculation yields that $\tau
_{12\cdots n}^{(1)}=1$ for the $n$-qubit state $GHZ$ and $\tau _{12\cdots
n}^{(1)}=0$ for the $n$-qubit state $W$.

\section{The $n$-tangle of odd $n$ qubits is an entanglement monotone}

Let $|\psi ^{\prime }\rangle $\ be also any state of $n$ qubits and $|\psi
^{\prime }\rangle =\sum_{i=0}^{2^{n}-1}b_{i}|i\rangle $, where $%
\sum_{i=0}^{2^{n}-1}|b_{i}|^{2}=1$. Two states $|\psi \rangle $ and $|\psi
^{\prime }\rangle $ are SLOCC\ entanglement\ equivalent if and only if there
exist invertible local operators $\mathcal{\alpha },\mathcal{\beta },\cdots $
such that \cite{Dur} 
\begin{equation}
|\psi ^{\prime }\rangle =\underbrace{\mathcal{\alpha }\otimes \mathcal{\beta 
}\otimes \cdots }_{n}|\psi \rangle .  \label{odd-1}
\end{equation}

The residual entanglement of odd $n$ qubits for the state $|\psi \rangle $
is defined as follows \cite{LDF07a}: 
\begin{equation}
\tau (\psi )=4\bigl|(\overline{\mathcal{I}}(a,n))^{2}-4\mathcal{I}^{\ast
}(a,n-1)\mathcal{I}_{+2^{n-1}}^{\ast }(a,n-1)\bigr|,
\label{odd-residual-def}
\end{equation}
where (see \cite{LDF07a,LDF09b}) 
\begin{align}
\overline{\mathcal{I}}(a,n) &=\sum_{i=0}^{2^{n-3}-1}(-1)^{N(i)} \Bigl[\bigl( %
a_{2i}a_{(2^{n}-1)-2i}-a_{2i+1}a_{(2^{n}-2)-2i}\bigr)  \notag \\
&\quad -\bigl( %
a_{(2^{n-1}-2)-2i}a_{(2^{n-1}+1)+2i}-a_{(2^{n-1}-1)-2i}a_{2^{n-1}+2i} \bigr) %
\Bigr],  \label{odd-2}
\end{align}
and (see \cite{LDF07a,LDF09b}) 
\begin{align}
\mathcal{I}^{\ast }(a,n-1)&=\sum_{i=0}^{2^{n-3}-1}(-1)^{N(i)} \bigl( %
a_{2i}a_{(2^{n-1}-1)-2i}-a_{2i+1}a_{(2^{n-1}-2)-2i}\bigr),  \label{odd-3} \\
\mathcal{I}_{+2^{n-1}}^{\ast }(a,n-1)&=\sum_{i=0}^{2^{n-3}-1}(-1)^{N(i)} %
\bigl(a_{2^{n-1}+2i}a_{(2^{n}-1)-2i}-a_{2^{n-1}+1+2i}a_{(2^{n}-2)-2i}\bigr).
\label{odd-4}
\end{align}

It has been also proven that if states $|\psi \rangle $ and $|\psi ^{\prime
}\rangle $ are SLOCC equivalent, then the following SLOCC equation holds 
\cite{LDF07a}: 
\begin{equation}
\tau (\psi ^{\prime })=\tau (\psi )\underbrace{\bigl|\det (\alpha )\det
(\beta )\det (\gamma )\cdots \bigr|^{2}}_{n}.  \label{odd-residual-2}
\end{equation}
\ 

We now argue that $\tau _{12\cdots n}^{(1)}=\tau (\psi )$. This can be seen
as follows. A simple calculation shows that $\overline{\mathcal{I}}(a,n)=T$
(see (i) in Appendix A). Inspection of Eqs. (\ref{n-tangle-10}) and (\ref%
{reduc-4}) (the reduced form of Eq. (\ref{odd-3})) reveals that $\mathcal{I}%
^{\ast }(a,n-1)=P/2$. Furthermore, inspection of Eqs. (\ref{n-tangle-11})
and (\ref{reduc-5}) (the reduced form of Eq. (\ref{odd-4})) reveals that $%
\mathcal{I}_{+2^{n-1}}^{\ast }(a,n-1)=Q/2$. Substituting these results into
Eq. (\ref{odd-residual-def}) yields 
\begin{equation}
\tau (\psi )=4\bigl|T^{2}-PQ\bigr|.  \label{transp-1}
\end{equation}%
Therefore, 
\begin{equation}
\tau _{12\cdots n}^{(1)}=\tau (\psi ).  \label{transp-2}
\end{equation}

Next we recall that the residual entanglement with respect to qubit $i$ is
defined as (see \cite{LDF09b}) $\tau ^{(i)}(\psi )$, which is obtained from $%
\tau (\psi )$ under the transposition $(1,i)$ on qubits 1 and $i$. The odd $%
n $-tangle is defined by taking the average of the residual entanglement
with respect to qubit $i$ \cite{LDF09b}: 
\begin{equation}
R(\psi )=\frac{1}{n}\sum_{i=1}^{n}\tau ^{(i)}(\psi ).  \label{residual-1}
\end{equation}%
Note that $R(\psi )$ is considered as an entanglement measure for odd $n$
qubits \cite{LDF09b}.

It follows immediately from Eq. (\ref{transp-2}) and the definitions of $%
\tau _{12\cdots n}^{(i)}$ and $\tau ^{(i)}(\psi )$ that 
\begin{equation}
\tau _{12\cdots n}^{(i)}=\tau ^{(i)}(\psi ),\quad i=1,2,\cdots ,n.
\label{transp-3}
\end{equation}

Further, Eq. (\ref{odd-tangle-3}), together with Eqs. (\ref{residual-1}) and
(\ref{transp-3}), yields 
\begin{equation}
\tau _{12\cdots n}=R(\psi ).  \label{tang-resid}
\end{equation}

A direct consequence of Eqs. (\ref{transp-3}) and (\ref{tang-resid}) is that
the $n$-tangle with respect to qubit $i$ and the $n$-tangle of odd $n$
qubits inherit the properties of the residual entanglement with respect to
qubit $i$ and the odd $n$-tangle. We highlight that the $n$-tangle with
respect to qubit $i$ and the $n$-tangle of odd $n$ qubits are $SL$-invariant
and $LU$-invariant, and are entanglement monotones (see \cite{LDF09b} for
details).

Clearly, both $\tau _{12\cdots n}^{(i)}$ and $\tau _{12\cdots n}$ satisfy
Eq. (\ref{odd-residual-2}). The $n$-tangle with respect to qubit $i$ is
called a SLOCC polynomial of degree 4 of odd $n$ qubits. It should be noted
that there are no polynomial invariants of degree 2 for odd $n$ qubits \cite%
{Luque}. In view of the SLOCC equation (\ref{odd-residual-2}), it is easy to
see that if one of $\tau _{12\cdots n}^{(i)}(\psi ^{\prime })$ (resp.  $\tau
_{12\cdots n}(\psi ^{\prime })$) and $\tau _{12\cdots n}^{(i)}(\psi )$
(resp. $\tau _{12\cdots n}(\psi )$) vanishes while the other does not, then $%
|\psi \rangle $ and $|\psi ^{\prime }\rangle $ belong to different SLOCC
classes. This reveals that the $n$-tangle with respect to qubit $i$ and the $%
n$-tangle of odd $n$ qubits can be used for SLOCC classification.

We exemplify the results for the GHZ state and the W state. In our previous
work \cite{LDFEPL} it has been shown that $\tau (GHZ)=1$ and $\tau (W)=0$
for any $n$-qubit GHZ and W states. The above analysis directly gives rise
to the conclusion that the $n$-tangle of odd $n$ qubits $\tau _{12\cdots n}$
is equal to 1 for the GHZ state and 0 for the W state.

Finally, we extend the $n$-tangle of odd $n$ qubits to mixed states via the
convex roof construction (see, e.g., the review \cite{Horodecki}): 
\begin{equation}
\tau_{12\cdots n}(\rho)=\min \sum_i p_i \tau _{12\cdots n}(\psi_i),
\end{equation}
where $p_i\ge 0$ and $\sum_i p_i=1$, and the minimum is taken over all
possible decompositions of $\rho$ into pure states, i.e. $\rho=\sum_i p_i
|\psi_i \rangle \langle\psi_i|$,

\section{Conclusion}

In summary, we have proposed the $n$-tangle of odd $n$ qubits, which is a
generalization of the standard form of the 3-tangle to any odd $n$-qubit
pure states. We have argued that the $n$-tangle of odd $n$ qubits is
invariant under permutations of the qubits, is an entanglement monotone. The 
$n$-tangle of odd $n$ qubits takes value 1 for the GHZ state and vanishes
for the W state. The $n$-tangle of odd $n$ qubits is considered as a natural
entanglement measure of any odd $n$-qubit pure states. Finally, we have
extended the $n$-tangle of odd $n$ qubits to mixed states via the convex
roof construction. Our results will provide more insight into the nature of
multipartite entanglement.

As is well known, two SLOCC inequivalent classes of three-qubit pure states,
namely the GHZ class and the W class, can be distinguished via the 3-tangle 
\cite{Dur,LDFPLA}. Polynomial invariants of degree 2 have been recently
exploited for SLOCC classification of four-qubit pure states \cite%
{LDF07b,LDFQIC09} and of the symmetric Dicke states with $l$ excitations of $%
n$ qubits \cite{LDFEPL}. More recently, four polynomial invariants of degree 
$2^{n/2}$ of any even $n$ qubits have been presented and several different
genuine entangled states inequivalent to the GHZ, the W, or the symmetric
Dicke states with $l$ excitations under SLOCC have been obtained by using
the polynomials \cite{LDFJPA}. Further attempts have been made to build
connections between polynomial (algebraic) invariants and SLOCC
classification \cite{Buniy,Viehmann}. We expect the $n$-tangle of odd $n$
qubits proposed in this paper can be used for SLOCC classification of any
odd $n$ qubits.

\section*{Appendix A}

\setcounter{equation}{0} \renewcommand{\theequation}{A\arabic{equation}}

We first give proofs of Eqs. (\ref{reduce-1})-(\ref{n-tangle-8}).

Let $\bar{\alpha}_{i}$ be the complement of $\alpha _{i}$. That is, $\bar{%
\alpha}_{i}=0$ when $\alpha _{i}=1$. Otherwise, $\bar{\alpha}_{i}=1$. In
view of that $\epsilon _{00}=\epsilon _{11}=0$ and $\epsilon _{01}=-\epsilon
_{10}=1$, to compute $W_{12\cdots n}^{(1)}$ in Eq. (\ref{odd-n-tangle-3}),
we only need to consider $\beta _{i}=\bar{\alpha}_{i}$, $\delta _{i}=\bar{%
\gamma}_{i}$, $i=2,\cdots ,n$, $\gamma _{1}={\bar\alpha}_{1}$, and $\delta
_{1}=\bar{\beta}_{1}$. Thus, Eq. (\ref{odd-n-tangle-3}) becomes 
\begin{equation}
W_{12\cdots n}^{(1)}=\sum a_{\alpha _{1}\alpha _{2}\cdots \alpha
_{n}}a_{\beta _{1}{\bar \alpha}_{2}\cdots \bar{\alpha}_{n}}a_{\bar{\alpha}%
_{1}\gamma _{2}\cdots \gamma _{n}}a_{\bar{\beta}_{1}\bar{\gamma}_{2}\cdots 
\bar{\gamma}_{n}}\times \epsilon _{\alpha _{2}\bar{\alpha}_{2}}\cdots
\epsilon _{\alpha _{n}\bar{\alpha}_{n}}\epsilon _{\gamma _{2}\bar{\gamma}%
_{2}}\cdots \epsilon _{\gamma _{n}\bar{\gamma}_{n}}\epsilon _{\alpha _{1}%
\bar{\alpha}_{1}}\epsilon _{\beta _{1}\bar{\beta}_{1}}.  \label{n-tangle-2}
\end{equation}

We distinguish two cases.

\textsl{Case 1.} $\beta _{1}=\alpha _{1}$.

In this case, $\epsilon _{\alpha _{1}\bar{\alpha}_{1}}\epsilon _{\beta _{1}%
\bar{\beta}_{1}}=1$. Thus, from Eq. (\ref{n-tangle-2}), we have 
\begin{equation}
W_{12\cdots n}^{(1)}=\sum a_{\alpha _{1}\alpha _{2}\cdots\alpha _{n}} a_{
\alpha _{1}\bar{\alpha} _{2}\cdots\bar{\alpha}_{n}}a_{\bar{\alpha}_{1}
\gamma_{2}\cdots\gamma _{n}}a_{\bar{\alpha}_{1}\bar{\gamma}_{2}\cdots \bar{
\gamma}_{n}} \times \epsilon _{\alpha _{2}\bar{\alpha}_{2}}\cdots\epsilon
_{\alpha _{n}\bar{\alpha}_{n}}\epsilon _{\gamma _{2}\bar{\gamma}
_{2}}\cdots\epsilon _{\gamma _{n}\bar{\gamma}_{n}}.  \label{tang-1}
\end{equation}

Letting 
\begin{align}
P&=\sum_{\alpha _{2}\cdots\alpha _{n}}a_{0\alpha _{2}\cdots\alpha _{n}} a_{0 
\bar{\alpha}_{2}\cdots\bar{\alpha}_{n}}\times \epsilon _{\alpha _{2}\bar{
\alpha}_{2}}\cdots \epsilon _{\alpha _{n}\bar{\alpha}_{n}},  \label{tang-2}
\\
Q&=\sum_{\alpha _{2}\cdots\alpha _{n}}a_{1\alpha _{2}\cdots\alpha _{n}} a_{1 
\bar{\alpha}_{2}\cdots\bar{\alpha}_{n}}\times \epsilon _{\alpha _{2}\bar{
\alpha}_{2}}\cdots\epsilon _{\alpha _{n} \bar{\alpha}_{n}},  \label{tang-3}
\end{align}
yields 
\begin{equation}
W_{12\cdots n}^{(1)}=2PQ.  \label{n-tangle-3}
\end{equation}

\textsl{Case 2.} $\beta _{1}=\bar{\alpha}_{1}$.

In this case, $\epsilon _{\alpha _{1}\bar{\alpha}_{1}}\epsilon _{\beta _{1} 
\bar{\beta}_{1}}=-1$. Thus, from Eq. (\ref{n-tangle-2}), we have 
\begin{equation}
W_{12\cdots n}^{(1)}=-\sum a_{\alpha _{1}\alpha _{2}\cdots\alpha _{n}} a_{ 
\bar{\alpha}_{1}\bar{\alpha}_{2}\cdots\bar{\alpha}_{n}} a_{\bar{\alpha}
_{1}\gamma_{2}\cdots\gamma _{n}}a_{\alpha _{1} \bar{\gamma}_{2}\cdots\bar{
\gamma}_{n}} \mathcal{\times }\epsilon _{\alpha _{2}\bar{\alpha}
_{2}}\cdots\epsilon _{\alpha _{n}\bar{\alpha}_{n}}\epsilon _{\gamma _{2}\bar{
\gamma}_{2}}\cdots\epsilon _{\gamma _{n}\bar{\gamma}_{n}}.
\label{n-tangle-4}
\end{equation}

Let 
\begin{align}
T& =\sum a_{0\alpha _{2}\cdots \alpha _{n}}a_{1\bar{\alpha}_{2}\cdots \bar{%
\alpha}_{n}}\times \epsilon _{\alpha _{2}\bar{\alpha}_{2}}\cdots \epsilon
_{\alpha _{n}\bar{\alpha}_{n}},  \label{n-tangle-5} \\
S& =\sum a_{1\alpha _{2}\cdots \alpha _{n}}a_{0\bar{\alpha}_{2}\cdots \bar{%
\alpha}_{n}}\times \epsilon _{\alpha _{2}\bar{\alpha}_{2}}\cdots \epsilon
_{\alpha _{n}\bar{\alpha}_{n}}.
\end{align}%
From that $\epsilon _{01}=-\epsilon _{10}=1$,\ $\epsilon _{\alpha _{i}\bar{%
\alpha}_{i}}=-\epsilon _{\bar{\alpha}_{i}\alpha _{i}}$, and therefore 
\begin{equation}
S=\sum a_{0\bar{\alpha}_{2}\cdots \bar{\alpha}_{n}}a_{1\alpha _{2}\cdots
\alpha _{n}}\times \epsilon _{\bar{\alpha}_{2}\alpha _{2}}\cdots \epsilon _{%
\bar{\alpha}_{n}\alpha _{n}}=T.
\end{equation}

Hence 
\begin{equation}
W_{12\cdots n}^{(1)}=-2T^{2}.  \label{n-tangle-7}
\end{equation}

Eq. (\ref{n-tangle-7}), together with Eq. (\ref{n-tangle-3}), yields 
\begin{equation}
W_{12\cdots n}^{(1)}=2(PQ-T^{2})\text{.}  \label{odd-tan-1}
\end{equation}

Inserting Eq. (\ref{odd-tan-1}) into Eq. (\ref{odd-tangle-1}) leads to 
\begin{equation}
\tau _{12\cdots n}^{(1)}=4\bigl| T^{2}-PQ\bigr|.  \label{odd-tan-2}
\end{equation}

Next, let $\alpha _{2}\cdots \alpha _{n}\ $be the binary representation of $%
i $. Noting that $(-1)^{N(i)}=\epsilon _{\alpha _{2}\bar{\alpha}_{2}}\cdots
\epsilon _{\alpha _{n}\bar{\alpha}_{n}}$, we may rewrite $T$ as 
\begin{equation}
T=\sum_{i=0}^{2^{n-1}-1}(-1)^{N(i)}a_{i}a_{2^{n}-i-1}.  \label{odd-tan-3}
\end{equation}

(i). Proof of $T=\overline{\mathcal{I}}(a,n)$

Expanding Eq. (\ref{n-tangle-5}), we obtain 
\begin{align}
T& =\sum a_{0\alpha _{2}\cdots \alpha _{n-1}0}a_{1\bar{\alpha}_{2}\cdots 
\bar{\alpha}_{n-1}1}\times \epsilon _{\alpha _{2}\bar{\alpha}_{2}}\cdots
\epsilon _{\alpha _{n-1}\bar{\alpha}_{n-1}}  \notag \\
& \quad -\sum a_{0\alpha _{2}\cdots \alpha _{n-1}1}a_{1\bar{\alpha}%
_{2}\cdots \bar{\alpha}_{n-1}0}\times \epsilon _{\alpha _{2}\bar{\alpha}%
_{2}}\cdots \epsilon _{\alpha _{n-1}\bar{\alpha}_{n-1}}  \notag \\
& =\sum a_{00\alpha _{3}\cdots \alpha _{n-1}0}a_{11\bar{\alpha}_{3}\cdots 
\bar{\alpha}_{n-1}1}\times \epsilon _{\alpha _{3}\bar{\alpha}_{3}}\cdots
\epsilon _{\alpha _{n-1}\bar{\alpha}_{n-1}}  \notag \\
& \quad -\sum a_{01\alpha _{3}\cdots \alpha _{n-1}0}a_{10\bar{\alpha}%
_{3}\cdots \bar{\alpha}_{n-1}1}\times \epsilon _{\alpha _{3}\bar{\alpha}%
_{3}}\cdots \epsilon _{\alpha _{n-1}\bar{\alpha}_{n-1}}  \notag \\
& \quad -\sum a_{00\alpha _{3}\cdots \alpha _{n-1}1}a_{11\bar{\alpha}%
_{3}\cdots \bar{\alpha}_{n-1}0}\times \epsilon _{\alpha _{3}\bar{\alpha}%
_{3}}\cdots \epsilon _{\alpha _{n-1}\bar{\alpha}_{n-1}}  \notag \\
& \quad +\sum a_{01\alpha _{3}\cdots \alpha _{n-1}1}a_{10\bar{\alpha}%
_{3}\cdots \bar{\alpha}_{n-1}0}\times \epsilon _{\alpha _{3}\bar{\alpha}%
_{3}}\cdots \epsilon _{\alpha _{n-1}\bar{\alpha}_{n-1}}  \notag \\
& =\overline{\mathcal{I}}(a,n),
\end{align}%
where the third equality follows by letting $\alpha _{3}\cdots \alpha _{n-1}$
be the binary number of $i$ and noting that$\ (-1)^{N(i)}=(-1)^{N(\alpha
_{3}\cdots \alpha _{n-1})}=\epsilon _{\alpha _{3}\bar{\alpha}_{3}}\cdots
\epsilon _{\alpha _{n-1}\bar{\alpha}_{n-1}}$.

(ii). Reduction of $P$

Expanding Eq. (\ref{tang-2}), we obtain 
\begin{align}
P& =\sum a_{0\alpha _{2}\cdots \alpha _{n-1}0}a_{0\bar{\alpha}_{2}\cdots 
\bar{\alpha}_{n-1}1}\times \epsilon _{\alpha _{2}\bar{\alpha}_{2}}\cdots
\epsilon _{\alpha _{n-1}\bar{\alpha}_{n-1}}  \notag \\
& \quad -\sum a_{0\alpha _{2}\cdots \alpha _{n-1}1}a_{0\bar{\alpha}%
_{2}\cdots \bar{\alpha}_{n-1}0}\times \epsilon _{\alpha _{2}\bar{\alpha}%
_{2}}\cdots \epsilon _{\alpha _{n-1}\bar{\alpha}_{n-1}}  \notag \\
& =2\sum a_{0\alpha _{2}\cdots \alpha _{n-1}0}a_{0\bar{\alpha}_{2}\cdots 
\bar{\alpha}_{n-1}1}\times \epsilon _{\alpha _{2}\bar{\alpha}_{2}}\cdots
\epsilon _{\alpha _{n-1}\bar{\alpha}_{n-1}}  \notag \\
& =2\sum_{i=0}^{2^{n-2}-1}(-1)^{N(i)}a_{2i}a_{2^{n-1}-2i-1},  \label{reduc-2}
\end{align}%
where the second equality follows from 
\begin{align}
& \sum a_{0\alpha _{2}\cdots \alpha _{n-1}1}a_{0\bar{\alpha}_{2}\cdots \bar{%
\alpha}_{n-1}0}\times \epsilon _{\alpha _{2}\bar{\alpha}_{2}}\cdots \epsilon
_{\alpha _{n-1}\bar{\alpha}_{n-1}}  \notag \\
& =-\sum a_{0\bar{\alpha}_{2}\cdots \bar{\alpha}_{n-1}0}a_{0\alpha
_{2}\cdots \alpha _{n-1}1}\times \epsilon _{\bar{\alpha}_{2}\alpha
_{2}}\cdots \epsilon _{\bar{\alpha}_{n-1}\alpha _{n-1}} \\
& =-\sum a_{0\alpha _{2}\cdots \alpha _{n-1}0}a_{0\bar{\alpha}_{2}\cdots 
\bar{\alpha}_{n-1}1}\times \epsilon _{\alpha _{2}\bar{\alpha}_{2}}\cdots
\epsilon _{\alpha _{n-1}\bar{\alpha}_{n-1}},
\end{align}%
and the third equality follows by letting $\alpha _{2}\cdots \alpha _{n-1}$
be the binary number of $i$ and noting that $(-1)^{N(i)}=\epsilon _{\alpha
_{2}\bar{\alpha}_{2}}\cdots \epsilon _{\alpha _{n-1}\bar{\alpha}_{n-1}}$.

(iii). Reduction of $Q$

Eq. (\ref{tang-3}) gives, by analogy with Eq. (\ref{reduc-2}), 
\begin{align}
Q&=\sum a_{1\alpha _{2}\cdots\alpha _{n-1}0}a_{1\bar{\alpha}_{2}\cdots \bar{
\alpha}_{n-1}1}\times \epsilon _{\alpha _{2} \bar{\alpha}_{2}}\cdots\epsilon
_{\alpha _{n-1}\bar{\alpha}_{n-1}}  \notag \\
&\quad -\sum a_{1\alpha _{2}\cdots\alpha _{n-1}1}a_{1\bar{ \alpha}_{2}\cdots%
\bar{ \alpha}_{n-1}0}\times \epsilon _{\alpha _{2}\bar{\alpha}%
_{2}}\cdots\epsilon _{\alpha _{n-1}\bar{\alpha}_{n-1}}  \notag \\
&=2\sum a_{1\alpha _{2}\cdots\alpha _{n-1}0}a_{1\bar{\alpha}_{2}\cdots\bar{
\alpha}_{n-1}1}\times \epsilon _{\alpha _{2}\bar{\alpha}_{2}}\cdots\epsilon
_{\alpha _{n-1}\bar{\alpha}_{n-1}}  \notag \\
&=2\sum_{i=0}^{2^{n-2}-1}(-1)^{N(i)}a_{2^{n-1}+2i}a_{2^{n}-2i-1}.
\label{reduc-3}
\end{align}

(iv). Reduction of $\mathcal{I}^{\ast }(a,n-1)$

By Eq. (\ref{odd-3}), we have 
\begin{equation}
\mathcal{I}^{\ast
}(a,n-1)=\sum_{i=0}^{2^{n-3}-1}(-1)^{N(i)}a_{2i}a_{(2^{n-1}-1)-2i}-%
\sum_{i=0}^{2^{n-3}-1}(-1)^{N(i)}a_{2i+1}a_{(2^{n-1}-2)-2i}.
\end{equation}

Let $k=2^{n-2}-1-i$. Then $N(k)+N(i)=n-2$, and hence $%
(-1)^{N(i)}=-(-1)^{N(k)}$, and 
\begin{equation}
\sum_{i=0}^{2^{n-3}-1}(-1)^{N(i)}a_{2i+1}a_{(2^{n-1}-2)-2i}=-%
\sum_{k=2^{n-3}}^{2^{n-2}-1}(-1)^{N(k)}a_{2k}a_{(2^{n-1}-1)-2k}.
\end{equation}
This leads to 
\begin{equation}
\mathcal{I}^{\ast
}(a,n-1)=\sum_{i=0}^{2^{n-2}-1}(-1)^{N(i)}a_{2i}a_{(2^{n-1}-1)-2i}.
\label{reduc-4}
\end{equation}

(v). Reduction of $\mathcal{I}_{+2^{n-1}}^{\ast }(a,n-1)$

By Eq. (\ref{odd-4}), we have 
\begin{equation}
\mathcal{I}_{+2^{n-1}}^{\ast
}(a,n-1)=\sum_{i=0}^{2^{n-3}-1}(-1)^{N(i)}a_{2^{n-1}+2i}a_{(2^{n}-1)-2i}-%
\sum_{i=0}^{2^{n-3}-1}(-1)^{N(i)}a_{2^{n-1}+1+2i}a_{(2^{n}-2)-2i}.
\end{equation}

Letting $k=2^{n-2}-1-i$, we have 
\begin{equation}
\sum_{i=0}^{2^{n-3}-1}(-1)^{N(i)}a_{2^{n-1}+1+2i}a_{(2^{n}-2)-2i}=-%
\sum_{k=2^{n-3}}^{2^{n-2}-1}(-1)^{N(k)}a_{2^{n-1}+2k}a_{(2^{n}-1)-2k}.
\end{equation}
This leads to 
\begin{equation}
\mathcal{I}_{+2^{n-1}}^{\ast
}(a,n-1)=\sum_{i=0}^{2^{n-2}-1}(-1)^{N(i)}a_{2^{n-1}+2i}a_{(2^{n}-1)-2i}.
\label{reduc-5}
\end{equation}

\end{document}